\documentclass[prb,twocolumn,showpacs,preprintnumbers,amsmath,amssymb]{revtex4-1}

\usepackage{graphicx}
\usepackage{dcolumn}
\usepackage{bm}
\usepackage[abs]{overpic}
\def\up{\uparrow}
\def\down{\downarrow }
\def\Vec#1{\bm{#1}}
\def\sla#1{\rlap/#1}

\begin{document}


\title{
Rotational Isotropy Breaking as Proof for Spin-polarized Cooper Pairs
in the Topological Superconductor Cu$_x$Bi$_2$Se$_3$
  }

\author{Yuki Nagai}
\affiliation{CCSE, Japan  Atomic Energy Agency, 5-1-5 Kashiwanoha, Kashiwa, Chiba, 277-8587, Japan}
\author{Hiroki Nakamura}
\affiliation{CCSE, Japan  Atomic Energy Agency, 5-1-5 Kashiwanoha, Kashiwa, Chiba, 277-8587, Japan}
\author{Masahiko Machida}
\affiliation{CCSE, Japan  Atomic Energy Agency, 5-1-5 Kashiwanoha, Kashiwa, Chiba, 277-8587, Japan}

\date{\today}

\begin{abstract}

In a promising candidate of topological superconductors, Cu$_x$Bi$_{2}$Se$_{3}$, we propose a way 
to exclusively determine the pairing symmetry. The proposal 
suggests that the angle dependence of the thermal conductivity in the basal $ab$-plane shows 
a distinct strong anisotropy only when the pairing symmetry is an odd-parity spin-polarized triplet 
below the superconducting transition temperature ($T_{\rm c}$).
Such striking isotropy breaking below $T_{\rm c}$ is explicitly involved in Dirac formalism for 
superconductors, in which the spin-orbit coupling is essential. 
We classify possible gap functions based on the Dirac formalism and clarify an origin of the isotropy breaking.

\end{abstract}

\maketitle
\section{Introduction}
The discovery of topological insulator has received considerable attention
because of its topologically protected nature of the surface gapless state. 
The non-trivial topology of the bulk insulator wavefunction shows up in 
edge boundaries together with closing the insulating gap \cite{Kane,HK_RMP10,Bernevig,Fu,Moore, Konig,FuKane,Nishide,TSato,Kuroda,Shen2}.  
A typical topological insulator, Bi$_{2}$Se$_{3}$ is characterized by a non-trivial $Z_{2}$ topology 
on its bulk valence band, which brings about a helical edge state 
leading to spontaneous spin current. 
Theoretically, such a non-trivial feature can be also maintained by opening of 
superconducting gap. Presently, a quest for the so-called topological superconductor 
is one of the most exciting issues in condensed matter physics.

Very recently, owing to its close vicinity to $Z_2$ topological insulator Bi$_{2}$Se$_{3}$, 
Cu intercalated material, Cu$_{x}$Bi$_{2}$Se$_{3}$ has been regarded as  
a key compound to investigate the non-trivial topological superconductivity 
\cite{Hor_L10,Wray_NP10,MKR_L11,FB_L10}.  
Indeed, Cu$_{x}$Bi$_{2}$Se$_{3}$ is a carrier doped compound, whose superconducting transition occurs around 0.3K.
Soon after the discovery, several groups observed zero-bias conductance peaks (ZBCP's) 
by using the point contact spectroscopy.
Generally, ZBCP has been well-known as a signature of unconventional superconductivity such as 
$d$-wave or sign-reversing $s$-wave superconductivity\cite{TanakaKashiwaya,NagaiPRB}. 
On the other hand, the origin of ZBCP observed in Cu$_{x}$Bi$_{2}$Se$_{3}$ 
may be ascribed to topologically-protected gapless Majorana 
fermion at edges\cite{QiRev,Schny_B08,Sato_B10} as a signature of topological superconductor.  
Then, clear evidence for such a protected edge state is now in great demand. 

So far, a tremendous number of studies have supported that 
the ZBCP is originated from Andreev bound-states formed at the edge boundary in unconventional superconductors. 
The emergence of the bound states is deeply associated 
with the internal sign change in the unconventional Cooper pair. 
In the superconductor, Cu$_{x}$Bi$_{2}$Se$_{3}$, Sasaki {\it et al.} theoretically demonstrated 
how the Majorana fermion
brings about ZBCP's \cite{Sasaki}. 
They examined four types of superconducting gap functions selected
by the point-symmetry analysis and drew a theoretical remark that an odd-parity spin-triplet 
is the most-likely pairing symmetry to explain the observed ZBCP's. 
However, they could not exclude the other paring symmetries, since 
any other three possible odd-parity pairing, one of whose gap is full and two of whose gaps 
have point-nodes, can induce ZBCP's with tiny variation of parameters 
in the original tight-binding model. 
Such confusing ambiguities reflect that it is very hard to identify the paring symmetry 
only through the energy dependence of the density of states. 
In this paper, we therefore propose that angle dependence of thermal conductivity is 
a crucial probe to break the above controversy.

In the history of the quest for superconducting pairing symmetry, the thermal conductivity 
has been frequently employed as a tool to identify the superconducting gap structure. 
Its angle-dependence obtained by rotating the applied magnetic field 
allows to explore the variation in the gap amplitude implying 
the existence of gap nodes \cite{YMatsuda,Izawa}. 
On the other hand, in topological superconductors, we do not need the application 
of the magnetic field, 
since the expected 
Majorana bound-state at edges is protected by time-reversal symmetry.   
In this paper, we present that anisotropy of the thermal conductivity 
is an exclusive proof to identify the paring state in the superconductor Cu$_{x}$Bi$_{2}$Se$_{3}$. 
The odd-parity spin-polarized Cooper-pair breaks the horizontal-angle 
invariance of the quasi-particle eigen-states around $\Gamma$-point, while 
non-polarized ones preserve the rotational invariance.  
The consequence is demonstrated by numerical calculations on the quasi-particle thermal conductivity and 
supported by theoretical analysis on a low-energy Dirac formalism for the 
superconductivity.  
Our proposal does not at all depend on the material parameters of the tight-binding model 
for Cu$_{x}$Bi$_{2}$Se$_{3}$, since the consequence is mathematically inherent in the formalism 
as long as its effective low-energy model is given by the Dirac-type one.
\section{Model and Method}
\begin{figure}[htbp]
\!\includegraphics[width = 0.8 \columnwidth]{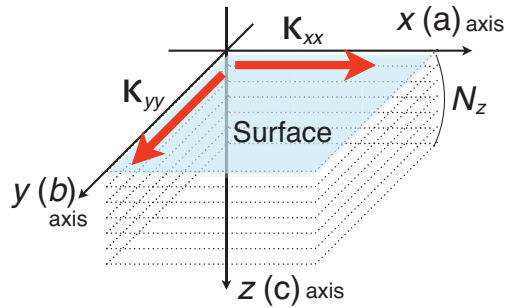} 
\caption{\label{fig:sites}
(Color online) The schematic figure of the calculation target system, which is discretized along $z$-direction. 
Then, the quasi-particles are localized around the edge surface plane at $z = 0$.%
}
\end{figure}

We start with a model Hamiltonian on the topological insulator Bi$_{2}$Se$_{3}$ 
proposed by several groups, which includes two-orbital spin-orbit coupling in $4 \times 4$ matrix \cite{Zhang,Sasaki}. 
In order to examine the edge state in the model with uniform superconducting gap, we have 
the mean-field Hamiltonian 
on discrete $N_{z}$ planes stacked along $z$-direction ($c$-axis)
based on the 
Bogoliubov-de Gennes formalism (see Fig.~\ref{fig:sites}.),
\begin{align}
H &= \sum_{k_{x},k_{y}} \sum_{i,j} c_{i}^{\dagger} {\cal H}_{ij}(k_{x},k_{y}) c_{j},
\end{align}
where $c_{i}$ is the $8$-component annihilation operator at the $i$-th plane, and  
$k_{x}$ and $k_{y}$ denote the in-plane momentum. 
Then, we have $8 \times 8 $ matrix Hamiltonian expressed as 
\begin{align}
{\cal H}_{ij}(k_{x},k_{y}) &=
\left(
\begin{array}{cc}
\hat \xi_{ij}(k_{x},k_{y}) & \hat{\Delta}\delta_{ij} \\
\hat{\Delta}^{\dagger}\delta_{ij} & -\hat \xi^{*}_{ij}(-k_{x},-k_{y})
\end{array}
\right),
\label{eq:Hamiltonian}
\end{align}
where, $\hat{\Delta}$ is $4 \times 4 $ matrix 
whose elements are given as 
$\Delta_{\sigma \sigma'}^{lm}$ using the orbital $l(m)$ and spin $\sigma(\sigma')$ indices. 
The normal-state Hamiltonian $\hat \xi(k_{x},k_{y})$ is written as 
\begin{align}
\hat \xi_{ij}(k_{x},k_{y}) &= \left(\begin{array}{cc}
\varepsilon^{+}_{ij}(k_{x},k_{y}) \hat{s}_{0} & \hat{A}_{ij}(k_{x},k_{y})  \\
 \hat{A}^{\dagger}_{ij}(k_{x},k_{y}) & \varepsilon^{-}_{ij}(k_{x},k_{y}) \hat{s}_{0}  
\end{array}\right),
\end{align}
where, $\hat{s}_{0}$ is the unit matrix in spin space, and 
$\varepsilon^{\pm}_{ij}(k_{x},k_{y})$ is expressed as 
$\varepsilon^{\pm}_{ij}(k_{x},k_{y}) = E_{\pm}(k_{x},k_{y}) \delta_{ij} + (-\bar{D}_{1}+\bar{B}_{1}) (\delta_{i j+1} + \delta_{i j-1})/2$. 
$E_{\pm} \equiv \epsilon(k_{x},k_{y}) \pm M(k_{x},k_{y})$, where $\epsilon(k_{x},k_{y}) \equiv 2 \bar{D}_{1} + 
\bar{D}_{2} \eta(k_{x},k_{y})  - \mu
$, $M(k_{x},k_{y}) \equiv M_{0} - 2 \bar{B}_{1} - \bar{B}_{2} \eta(k_{x},k_{y}) $ and $\eta(k_{x},k_{y}) \equiv (4/3)(3 - 2 \cos(k_{x} \sqrt{3}/2)) \cos(k_{y}/2 - \cos k_{y})$. 
The $2  \times 2 $ matrix $\hat{A}_{ij}(k_{x},k_{y})$ is given by 
\begin{align}
\hat{A}_{ij}(k_{x},k_{y}) &= \left(\begin{array}{cc} 
\bar{A}_{1} \hat{R}^{s}_{ij}& A_{2}^{-}(k_{x},k_{y}) \delta_{ij} \\
A_{2}^{+}(k_{x},k_{y})\delta_{ij}  & -\bar{A}_{1} \hat{R}^{s}_{ij}
\end{array}\right),
\end{align}
with the matrix $\hat{R}_{s}$ whose elements 
is expressed by $[\hat{R}_{s}]_{ij} = -i \delta_{i j+1}/2+ i \delta_{i j-1}/2$ generated by Fourier transformation of $\sin k_{z}$ and $A_{2}^{\pm}(k_{x},k_{y}) \equiv (2/3)\bar{A}_{2} (\sqrt{3} \sin(k_{x} \sqrt{3}/2) \cos(k_{y}/2) + \pm i (\cos(k_{x}\sqrt{3}/2) \sin( k_{y}/2) + \sin(k_{y})))$. 
We set 
$M_{0} = 0.28$ eV, $\mu = 0.5$ eV, $\bar{A}_{1} = 0.32$ eV, $\bar{A}_{2} = 4.1/a$ eV, $\bar{B}_{1} = 0.216$ eV, $\bar{B}_{2} = 56.6/a^{2}$ eV, $\bar{D}_{1} = 0.024$ eV, $\bar{D}_{2} = 19.6/a^{2}$ and $a =  4.076$ \AA  \: as the material parameters for Cu$_{x}$Bi$_{2}$Se$_{3}$\cite{Sasaki}. 
According to Refs.~\cite{Fu,Sasaki,Hao}, we examine four different types of the superconducting gap function, $\Delta_{1}$ to $\Delta_{4}$, which cover possible all gap functions 
selected by the point-group symmetry analysis (see also Table \ref{tb:tb1}). 

In order to obtain the angle dependence of the thermal-conductivity, we calculate the 
quasi-particle thermal conductivity tensor $\kappa_{ij}(T)$ expressed as\cite{Bannemann,Fisk} 
\begin{align}
\kappa_{ij} &= \frac{1}{T} \sum_{k,n} v_{i}^{n}(k) v_{j}^{n}(k) \tau(E_{k}^{n}) E_{k}^{n 2} \left( -\frac{\partial f}{\partial E_{k}^{n}} \right), \label{eq:kappa}
\end{align}
in which $v_{i}^{n}(k)$, $E_{k}^{n}$, and $\tau(E_{k}^{n})$, respectively, are quasi-particle velocity, energy, and relaxation time as a function 
of $k$ and the band index $n$, and $f$ is Fermi distribution function. 
For simplicity, we take $\tau(E_{k}) = \tau$. 
Then, we calculate the thermal conductivity tensor Eq.~(\ref{eq:kappa}) by diagonalizing 
$8 N_{z} \times 8 N_{z}$ matrix BdG Hamiltonian (\ref{eq:Hamiltonian}). 
It is noted that quasi-particles are localized around 
the top edge surface 
as shown in Fig.~\ref{fig:sites}.  
We emphasize that 
the calculation result $\kappa_{ij}$ does not qualitatively change, even if one 
takes into account the other contributions to 
the thermal conductivity. 
This is because a key feature is whether the quasi-particle eigen-state, i.e., the electronic structure is 
horizontally angle dependent around $\Gamma$-point or not. 

\begin{figure}[htbp]
\begin{center}
\!\includegraphics[width = 0.9 \columnwidth]{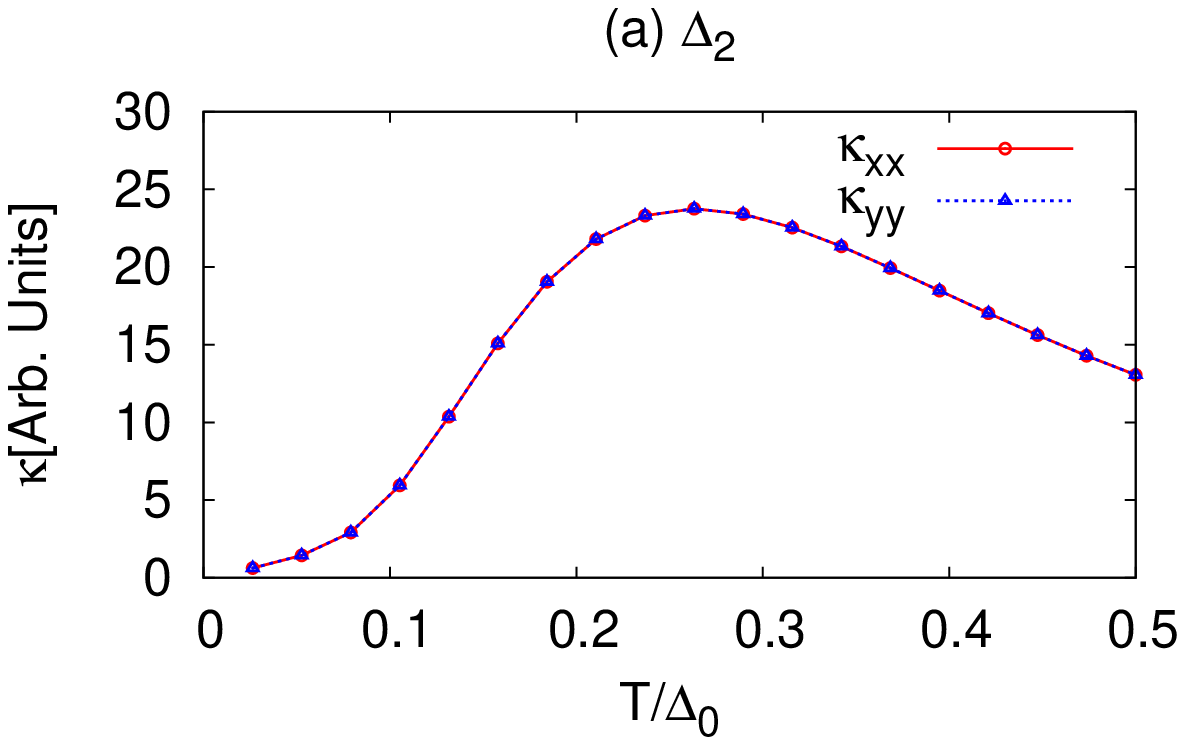} \\
\includegraphics[width = 0.9 \columnwidth]{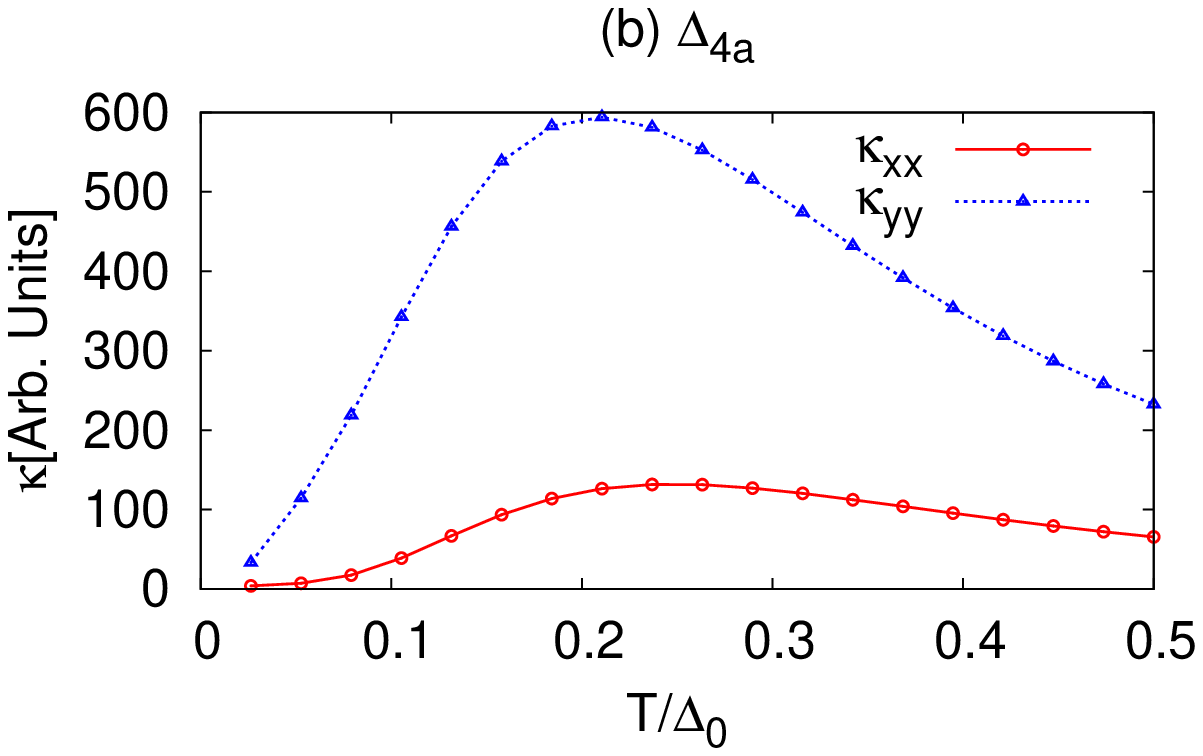}
\caption{\label{fig:kappa}
(Color online) The temperature dependence of the thermal conductivity carried by the edge bound states for  
the gap functions (a) $\Delta_{2}$ and (b) $\Delta_{4a}$. $N_{z} = 64$.
}
\end{center}
\end{figure}
\section{Numerical Results}
\begin{figure*}[htbp]
 \begin{center}
     \begin{tabular}{p{1 \columnwidth} p{1\columnwidth}}
      \resizebox{0.9 \columnwidth}{!}{\includegraphics{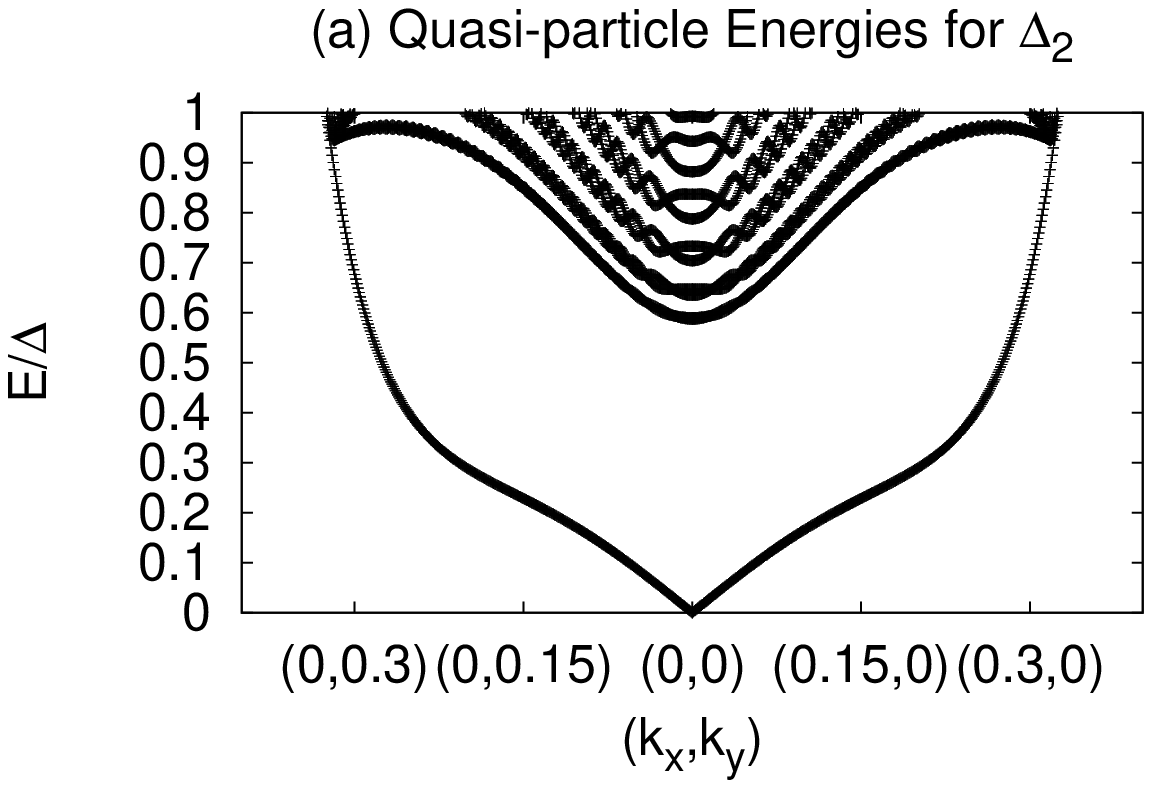}} &
      \resizebox{0.65 \columnwidth}{!}{\includegraphics{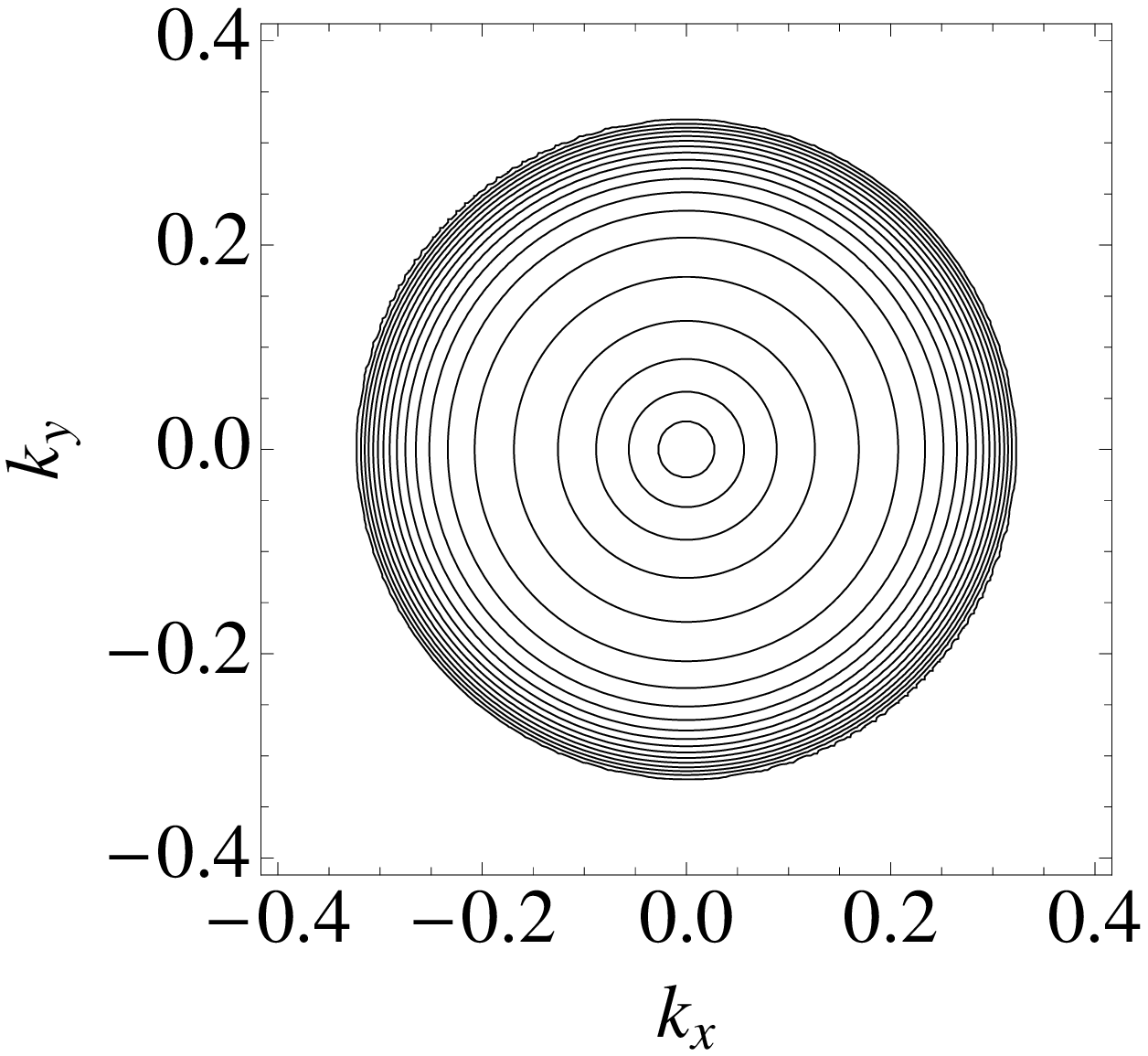}} \\
      \resizebox{0.9 \columnwidth}{!}{\includegraphics{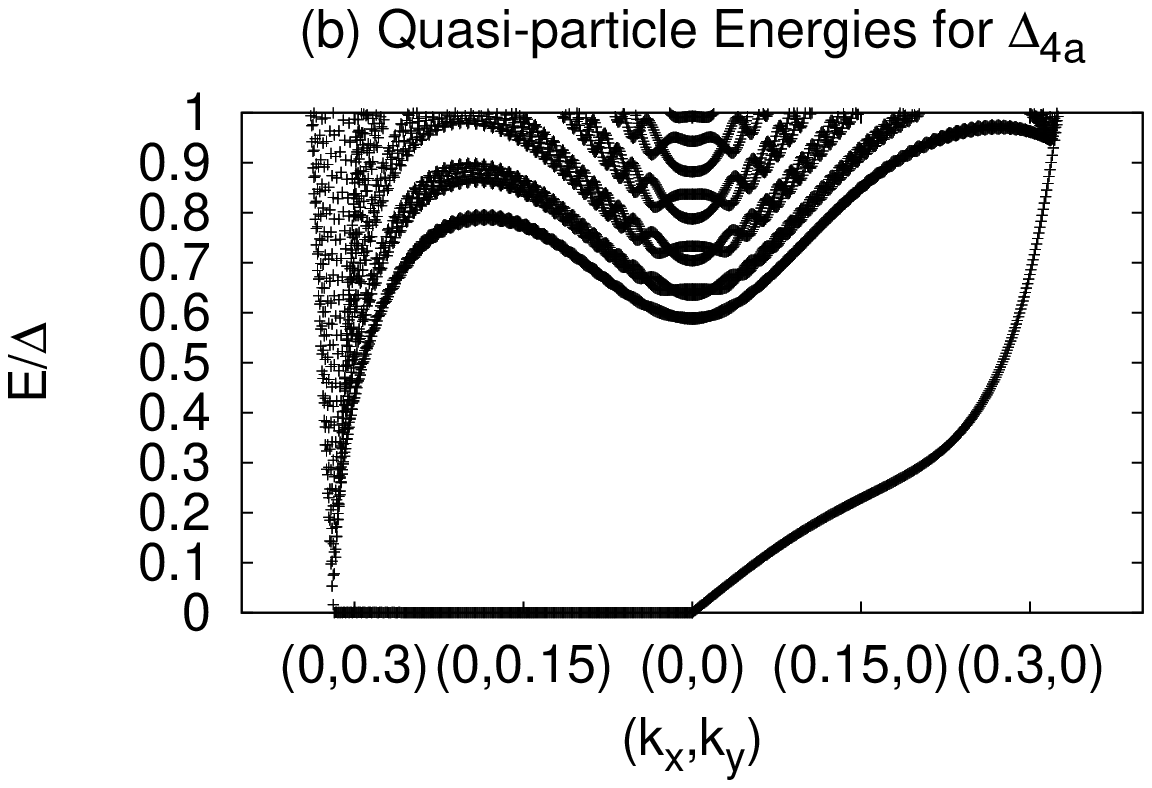}} &
      \resizebox{0.65 \columnwidth}{!}{\includegraphics{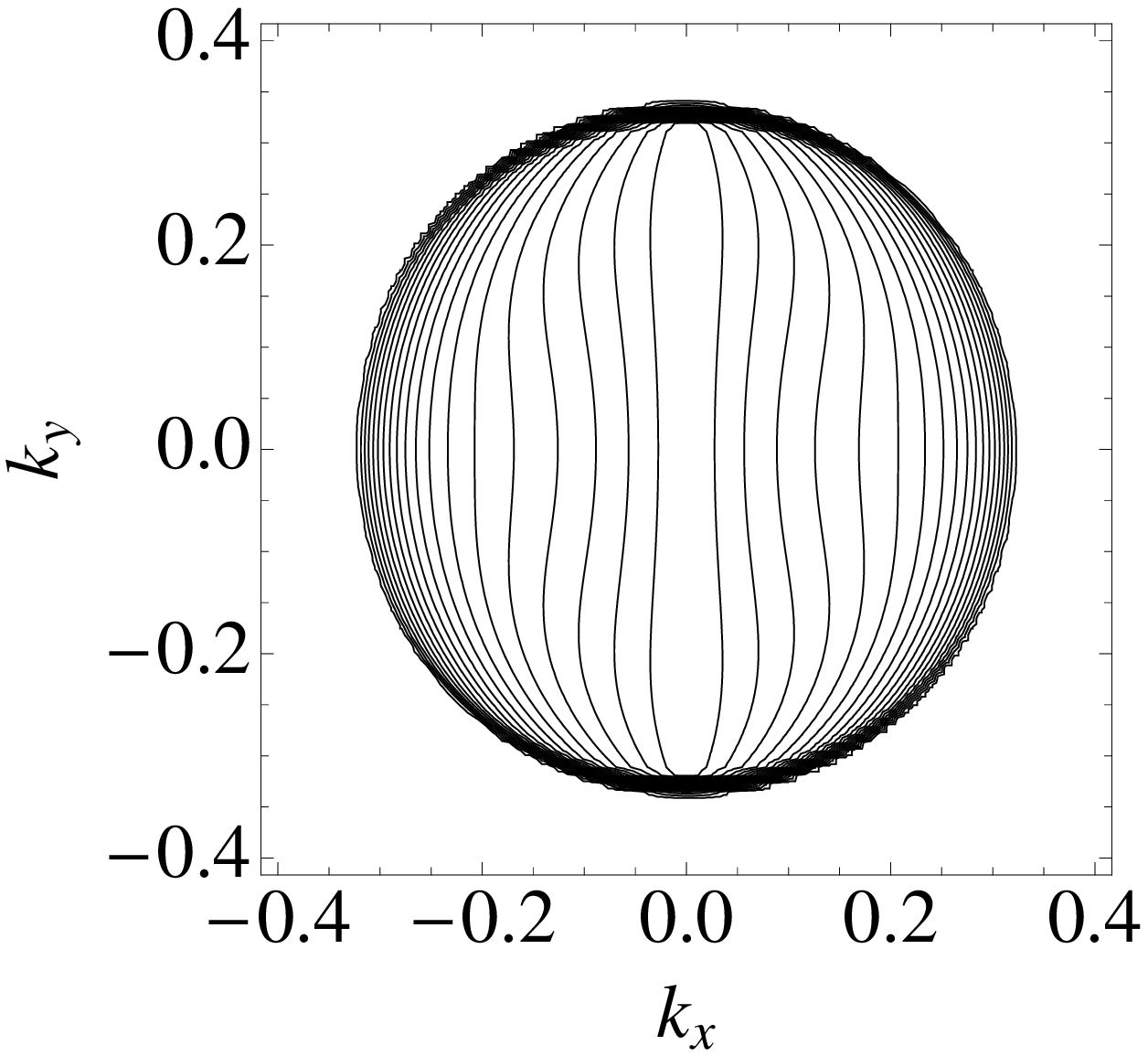}} 
    \end{tabular}
\caption{\label{fig:eigen}
(Left) The distributions of quasi-particle energies in in-plane momentum space with $N_{z} = 64$ for the gap functions 
(a) $\Delta_{2}$ and (b) $\Delta_{4a}$, respectively. 
(Right) The contour plots of the minimum eigenvalues regarded as the edge states for 
(a) $\Delta_{2}$ and (b) $\Delta_{4a}$, respectively. 
 The contours are drawn at regular intervals ($\delta \Delta = \Delta_{0}/20$). }
 \end{center}
\end{figure*}

In order to examine whether the isotropy of $\kappa_{ij}$ is preserved or not, we obtain the temperature dependence of $\kappa_{xx}$ and $\kappa_{yy}$ well below $T_{\rm c}$. 
Here, we drop 
the temperature-dependence of the pair amplitude as 
$\Delta(T) \propto \Delta_{0} = 0.05$ eV for convenience of calculations. 
Then, the result is valid in the low temperature range 
as shown in Fig.~\ref{fig:kappa}.
%
The thermal conductivity from the bulk body mainly arising from phonon is always isotropic free from the present argument. 
As shown in Fig.~\ref{fig:kappa},  the thermal conductivity is 
found to be isotropic and anisotropic reflecting the difference in the gap function. 
For the gap function $\Delta_{4a}$ ($\Delta_{\up \up}^{12} = \Delta_{\down \down}^{12} = - \Delta_{\up \up}^{21}
 = -\Delta_{\down \down}^{21}$. See, Table ~\ref{tb:tb1}), one finds a clear anisotropy, while not for $\Delta_{2}$. 
This difference is explained by 
the horizontal-angle dependence of diagonalized (edge-state's) eigenvalue distribution as shown in Fig.~\ref{fig:eigen}.  
These results indicate that the gap function $\Delta_{4a}$ breaks the rotational isotropy in 
the original model without the gap function while other ones do not show any isotropy breaking. 

\section{Origin of the Rotational Isotropy Breaking: Massive Dirac Equations}
Now, let us theoretically pursue the origin of the above result.
First, using a Nambu space not commonly used in condensed matter theories, 
we introduce the massive Dirac Hamiltonian with superconducting pair function (see Appendix A.)\cite{note,Peskin,Shovkovy}, 
\begin{align}
H =\int d{\bm r} \left(\begin{array}{cc}\bar{\psi}({\bm r}) & \bar{\psi}_{c}({\bm r})  \end{array}\right)
 \left(\begin{array}{cc}h_{-}({\bm r}) & \Delta^{-}({\bm r})  \\
 \Delta^{+}({\bm r}) & h_{+}({\bm r})
\end{array}\right) \left(\begin{array}{c}\psi({\bm r}) \\
\psi_{c}({\bm r})
\end{array}\right), \label{eq:massiveH}
\end{align}
with 
\begin{align}
h_{\pm} &= M_{0} -
 \gamma^{1} i \partial_{x} -
     \gamma^{2}i \partial_{y} -   
  \gamma^{3} i \partial_{z}  \pm \mu \gamma^{0}.
\end{align}
Here, $\Delta^{-} \equiv \gamma^{0} \hat{\Delta} i \gamma^{2}$, $\Delta^{+} \equiv \gamma^{0} (\Delta^{-})^{\dagger} \gamma^{0}$, and 
$\gamma^{i}$ is $4 \times 4$ Dirac gamma matrix in the Dirac representation. 
$\bar{\psi}(x) \equiv \psi^{\dagger}(x)\gamma^{0}$, $\bar{\psi}_{c}(x) \equiv \psi_{c}^{\dagger} \gamma^{0}$, $\psi_{c} \equiv {\cal C} \bar{\psi}^{T}$, where ${\cal C} (\equiv i \gamma^{2} \gamma^{0})$ is a representative matrix of charge-conjugation \cite{Nambu}. 
The space is 
three-dimensional (3-D) $x$-$y$-$z$ coordinate system.
Utilizing $2 \times 2$ Pauli matrices $\hat{\sigma}_{i}$ in the orbital space and $\hat{s}_{i}$ in the spin space, 
gamma matrices are represented as $\gamma^{0} = \hat{\sigma}_{z} \otimes 1$, $\gamma^{i = 1,2,3} = i \hat{\sigma}_{y} \otimes \hat{s}_{i}$ and $\gamma^{5} = \hat{\sigma}_{x} \otimes 1$, respectively,  with 
the relation $\gamma^{5} = i \gamma^{0} \gamma^{1} \gamma^{2} \gamma^{3}$.  
We note that parity of $\gamma^{\mu \neq 0}$ is odd since the transformation property is 
equivalent with that of vector. 

We classify the possible gap functions with gamma matrices in Table \ref{tb:tb1}. 
Reminding the symmetry of $\hat{\Delta}({\bm k}) = \hat{\Delta}(-{\bm k})$, we have 
only six $\hat{\Delta}$ matrices constrained by the fermion anti-symmetric relation $- \hat{\Delta} = \hat{\Delta}^{T}$,  
\begin{align}
\hat{\Delta} &\propto  {\cal C}, {\cal C} \gamma^{5}, {\cal C}  \gamma^{\mu} \gamma^{5},
\end{align}
with $\mu = 0,1,2,3$. 
According to 
Table \ref{tb:tb1} (a list of $\hat{\Delta}$), 
the gap functions $\hat{\Delta}_n^{-}$ introduced by Fu and Berg\cite{Fu,Hao} 
are characterized by a scalar, a pseudo-scalar, and a four-vector: 
\begin{align}
\Delta^{-}_{n} &\propto 
\left\{ \begin{array}{ll}
1  & (n = {\rm 2}) \\
 \gamma^{5} & (n = {\rm 1a}) \\
\sla{\alpha} \gamma^{5} & ({\rm else}) \\
\end{array} \right. ,
\end{align}
where, the Feynman slash $\sla{\alpha}$ is defined by $\sum_{\mu} \gamma^{\mu }\alpha_{\mu}$, and 
the gap function is characterized by the four-vector $\alpha$. 
For example, the gap function $\Delta_{4a}$ is characterized by $\alpha_{\mu} = \delta_{\mu 2}$. 
The $s$-wave pairing gap function $\Delta_{1a}$ in this representation has been well-studied 
in color superconductivity\cite{Shovkovy} .
From the above compact expressions for the gap functions, 
one finds that 
the present Nambu representation using the charge-conjugation operator 
is quite convenient in the massive Dirac equations 
for the superconductivity.  
In terms of this representation, we also show that gap functions represented by $\sla{\alpha}$ with $\gamma^{\mu \neq 0}$-components have point-nodes and know the positions of these point-nodes (see Appendix B). 

In the normal state, the Hamiltonian is rotationally invariant in 2-D $xy$-space. 
On the other hand, in the superconducting state, the gap function represented by $\sla{\alpha}$ with $\gamma^{\mu \neq 0}$-components 
breaks the rotational symmetry, indicating non-zero spatial elements of four-vector $\alpha$.  
Thus, the diagonalization together with $\hat{\Delta}$ breaks the isotropy of quasi-particles. 
Of course, 
one easily finds, {\it before the diagonalization step}, that the gap function $\Delta_{2}$ 
maintains isotropy and full gap 
while
$\Delta_{4}$ does not. 
Thus, as shown in Fig.~\ref{fig:eigen}, 
the distribution of eigenvalues of Eq.~(\ref{eq:Hamiltonian}) is horizontally angle-dependent only 
in the case of $\hat{\Delta}_{4}$. 

%
%
%
\begin{table*}[htdp]
\begin{center}\begin{tabular}{|c|c|c|c|c|c|}
\hline  &
$ \Delta$&
$\hat{\Delta}_{n}$  &
Parity  &
  $\hat{\Delta}^{-}$ &
   $\bar{\psi} \hat{\Delta}^{-} \psi_{c}$ \\
 \hline 
Intra-orbital singlet: $\hat{\sigma}_{0} \otimes i \hat{s}_{y}  \Delta_{0}$
  &
 ${\cal C} \gamma^{5}$&
  $\hat{\Delta}_{1a}$  &
   $+$ &
  $\gamma^{5}$&
   Scalar \\ 
      \hline 
Intra-orbital singlet: $\hat{\sigma}_{z} \otimes i \hat{s}_{y}  \Delta_{0}$
  &
    ${\cal C}  \gamma^{0} \gamma^{5}$ &
  $\hat{\Delta}_{1b}$  &
   $+$ &
  $\gamma^{0} \gamma^{5}$&
   $t$-Polar \\ 
     \hline 
Inter-orbital singlet: $\hat{\sigma}_{x} \otimes i \hat{s}_{y} \Delta_{0}$  &
 ${\cal C}$ &
  $\hat{\Delta}_{2}$  &
   $-$ &
  $1$&
   P-Scalar \\ 

         \hline 
Inter-orbital triplet: $\hat{\sigma}_{x} \otimes i (\Vec{d}^{x} \cdot \hat{\Vec{s}}) \hat{s}_{y}$
  &
     ${\cal C}  \gamma^{1} \gamma^{5}$ &
  $\hat{\Delta}_{4b}$  &
   $-$ &
  $\gamma^{1} \gamma^{5}$&
   $x$-Polar \\ 
        \hline 
   Inter-orbital triplet: $\hat{\sigma}_{x} \otimes i (\Vec{d}^{y} \cdot \hat{\Vec{s}}) \hat{s}_{y}$       
  &        
        ${\cal C}  \gamma^{2} \gamma^{5}$ &
  $\hat{\Delta}_{4a}$  &
   $-$ &
  $\gamma^{2} \gamma^{5}$&
   $y$-Polar \\ 
           \hline 
             Inter-orbital triplet: $\hat{\sigma}_{x} \otimes i (\Vec{d}^{z} \cdot \hat{\Vec{s}}) \hat{s}_{y}$ 
  &${\cal C}  \gamma^{3} \gamma^{5}$ &
  $\hat{\Delta}_{3}$  &
   $-$ &
  $\gamma^{3} \gamma^{5}$&
   $z$-Polar \\ 
\hline
\end{tabular} \caption{
The representation of the gap functions with gamma matrices. See 
text for $\hat{\Delta}_{n}$. 
``P-Scalar'' denotes a pseudo scalar whose parity is odd 
and ``$i$-Polar'' denotes a polar vector in $i$-direction in four-dimensional space. 
$\Vec{d}^{\nu}$ denotes a $\Vec{d}$-vector in the $\nu$-direction ($[\Vec{d}^{\nu}]_{i} = \Delta_{0} \delta_{i \nu}$). 
$\hat{\sigma}_{i}$ and $\hat{s}_{i}$ denote $2 \times 2$ Pauli matrices in the orbital space and the spin space, respectively.}
\label{tb:tb1}
\end{center}
\label{defaulttable}
\end{table*}
%

\section{Discussion}
Finally, let us discuss the feasibility of the anisotropic thermal conductivity as a probe of pairing state. 
The first issue is quantitative predictability.
In the present theoretical treatment, 
we neglect multi-orbital effects on quasi-particle scattering for simplicity 
 and handle only orbital-diagonal one. 
Then, it means that quantitative predictability of the thermal conductivity is beyond the present scope. 
On the other hand, the anisotropy caused by 
the gap function $\Delta_{4}$ is still robust, since the quasi-velocities $v_{x}$ and $v_{y}$ 
 become anisotropic below $T_{\rm c}$ owing to the coupling with $\Delta_{4}$. 
Clearly, such a contribution is  a leading one in sufficiently low-temperature range $T/\Delta_{0} < 1$ as shown in Fig.~\ref{fig:kappa}, where 
the bulk thermal conductivity 
irrelevant to 
the present 
mechanism is significantly reduced  because of the superconducting gap opening.
The next is qualitative comparison with other gap cases. 
In non-topological superconductivity 
$\Delta_{1a}$ ($\Delta_{\up \down}^{11} = - \Delta_{\down \up}^{11} = \Delta_{\up \down}^{22} = -\Delta_{\down \up}^{22}$) 
the thermal conductivity is rather small and isotropic because of no edge bound state in contrast other gaps. 
For $\Delta_{2}$ and $\Delta_{3}$ ($\Delta_{\up \down}^{12} = \Delta_{\down \up}^{12} = -\Delta_{\up \down}^{21} 
= - \Delta_{\down \up}^{21}$), the gapless edge bound states might appear 
depending on the band parameters, but the contribution to the thermal conductivity is isotropic since these gap functions are symmetric around $\Gamma$-point in $k_{x}$-$k_{y}$ plane. 
Thus, we point out that the isotropy breaking of the thermal conductivity is sufficiently
an exclusive evidence of the spin-polarized Cooper pair ($\Delta_{4}$). 

Moreover, our proposal does not depend on the material parameter values such as 
$M_{0}$, $\mu$, $\bar{A}_{1}$, $\bar{A}_{2}$, $\bar{B}_{1}$, $\bar{B}_{2}$, $\bar{D}_{1}$, and $\bar{D}_{2}$. 
On the other hand, it should be noted that non-electronic 
disorder or other scattering contributions might recover the original isotropy. 
However, the isotropy breaking mechanism is 
inherent in the topological superconductivity modeling. 
This idea is applicable for not only various topological superconductors 
but also superconductivity emerged in high-energy physics.

\section{Conclusion}
In conclusion, we numerically calculated the thermal conductivity tensor $\kappa_{xx}$ and $\kappa_{yy}$ dominated by 
the edge bound states in the superconductor 
Cu$_x$Bi$_{2}$Se$_{3}$. 
Consequently, we found that the rotational isotropy of the thermal conductivity is broken below 
$T_{\rm c}$ for the spin-polarized gap $\Delta_{4}$. 
In order to explore the mechanism, we noticed that the Nambu representation 
expanded by charge-conjugation operator is convenient 
in handling massive Dirac equations with Cooper pair functions and 
newly classified the possible gap functions by using simple mathematical tools 
as gamma matrices $\gamma$ and four-vector $\alpha$. 
As a result of the classification, we found that 
the horizontal angle invariance of 
the quasi-particle eigenvalues around $\Gamma$-point is broken for the spin-polarized 
pair $\Delta_{4}$, which is mathematically characterized by four-vector $\alpha$ 
lying on the basal xy-plane. We propose that the angle dependence of the thermal conductivity
 has an exclusive tool to identify the gap function in topological superconductors. 
\section*{Acknowledgment}
We thank M. Okumura, Y. Ota, and T. Koyama for helpful discussions and comments. 
The calculations have been performed using the supercomputing 
system PRIMERGY BX900 at the Japan Atomic Energy Agency. 
This research was partially supported by a Grant-in-Aid for Scientific Research from JSPS (Grant No. 24340079). 


\appendix
\section{Massive Dirac Hamiltonian}
We rewrite the BdG Hamiltonian to introduce the massive Dirac Hamiltonian with superconducting pair function (\ref{eq:massiveH}).
We start with the conventional BdG Hamiltonian expressed as 
\begin{widetext}
\begin{align}
H =\int d\Vec{r} \left(\begin{array}{cc}\psi^{\dagger}(\Vec{r}) & \psi(\Vec{r})^{T}  \end{array}\right)
 \left(\begin{array}{cc}h(\Vec{r}) -\mu& \Delta(\Vec{r})  \\
 \Delta^{\dagger}(\Vec{r}) & -h(\Vec{r})^{\ast}+\mu
\end{array}\right) \left(\begin{array}{c}\psi(\Vec{r}) \\
\psi^{\ast}(\Vec{r})
\end{array}\right). 
\end{align}
\end{widetext}
Here, the Hamiltonian in normal states is written as
\begin{align}
h(\Vec{r}) &= M_{0} \gamma^{0} -
 \gamma^{0} \gamma^{1} i \partial_{x} -
  \gamma^{0}   \gamma^{2}i \partial_{y} -   
 \gamma^{0}  \gamma^{3} i \partial_{z}. 
\end{align}
We introduce the charge-conjugation operator,
\begin{align}
\psi_{c}(\Vec{r}) &= {\cal C} \bar{\psi}^{T}(\Vec{r}) = {\cal C} \gamma^{0} \psi^{\ast}(\Vec{r}),
\end{align}
with the charge-conjugation matrix ${\cal C} \equiv i \gamma^{2} \gamma^{0}$ and $\bar{\psi}(\Vec{r}) \equiv \psi^{\dagger}(\Vec{r}) \gamma^{0}$. 
The gamma matrices have the anti-commutation relation:
\begin{align}
\gamma^{\mu} \gamma^{\nu} + \gamma^{\nu} \gamma^{\mu} = 2 \eta^{\mu \nu},
\end{align}
with $\eta = {\rm diag} \: (1,-1,-1,-1)$. 
We note that both the gamma matrices $\gamma^{\mu = 1,2,3}$ and ${\cal C}$ are the 
non-Hermitian matrices ($\gamma^{\mu = 1,2,3 \dagger} = - \gamma^{\mu=1,2,3}$ and ${\cal C}^{\dagger} = - {\cal C}$).
The matrix element $\psi^{T}(\Vec{r})(- h^{\ast}(\Vec{r}) ) \psi^{\ast}(\Vec{r})$ is rewritten as 
\begin{align}
\psi^{T}(\Vec{r}) (-h(\Vec{r})^{\ast}) \psi(\Vec{r})^{\ast}  &= \psi_{c}^{\dagger}(\Vec{r})\gamma^{0} {\cal C}^{\dagger} (-h(\Vec{r})^{\ast}) {\cal C} \gamma^{0} \psi_{c}(\Vec{r}), \nonumber \\
&= \bar{\psi}_{c}(\Vec{r}) h(\Vec{r}) \psi_{c}(\Vec{r}).
\end{align}
Therefore, we obtain the massive Dirac Hamiltonian with the charge-conjugation operators (\ref{eq:massiveH}).
\section{Positions of point-nodes}
In terms of the four vector $\Vec{\alpha}$, one can easily find the positions of point-nodes.
For example, 
the four vector for $\Delta_{4a}$ is characterized by $\Vec{\alpha} = (0,0,1,0)$. 
In this case, the gap function has the rotational symmetry around $y$-axis. 
Therefore, the point-nodes can exist only on the $k_{y}$-axis.
Generalizing the above, we can show 
that gap function represented by polar-vectors $\hat{\Delta}^{-} = \Delta_{0} \gamma^{\nu = 1,2,3} \gamma^{5}$ can have point-nodes.
As we discussed in the above paragraph, 
the point-nodes can exist on the $k_{\nu}$-axis, since 
the gap function represented by a $\nu$-polar vector has the rotational symmetry around $\nu$-axis. 
Thus, we only consider the case of $k_{\mu \neq \nu} = 0$. 
In this case, the normal-Hamiltonian in momentum space is written as 
\begin{align}
h(\Vec{k}) &= M_{0} \gamma^{0} + \delta_{\mu \nu }\gamma^{0} k_{\mu} \gamma^{\mu}.
\end{align}
The BdG equations are expressed as
\begin{align}
(\gamma^{0} M_{0} + \delta_{\mu \nu } \gamma^{0} k_{\mu} \gamma^{\mu} - \mu ) u + \gamma^{0} \gamma^{\nu} \gamma^{5} \Delta_{0} v &= E u, \\
\gamma^{0} \gamma^{\nu} \gamma^{5} \Delta_{0} u + (\gamma^{0} M_{0} + \gamma^{0} k_{\mu} \gamma^{\mu} + \mu ) v  &= E v.
\end{align}
If the point-nodes exist, the above equations have the zero-energy solutions ($E = 0$).
By solving the BdG equations, we obtain the relation,
\begin{align}
 k_{\nu}^{2} &= \mu^{2} - M_{0}^{2} + \Delta_{0}^{2}.
\end{align}
On the other hand, the Fermi surface in normal states is determined by the following relation,
\begin{align}
 k_{x}^{2} + k_{y}^{2} + k_{z}^{2} &= \mu^{2} - M_{0}^{2}.
\end{align}
Therefore, if the $\Delta_{0}$ is small, the point-nodes always exist near the normal-states Fermi surface.


\begin{thebibliography}{99}
\bibitem{note}
There are two ``Nambu space'', which are constructed by a conventional particle-hole spinor $(\psi,\psi^{\ast})$ and a particle-antiparticle spinor $(\psi,{\cal C} \bar{\psi}^{T})$.
\end{thebibliography}
\end{document}